\documentclass[aps]{revtex4}
\usepackage{graphicx,psfrag}
\def\bc{\begin{center}}
\def\ec{\end{center}}

\def\beq{\begin{equation}}
\def\eeq{\end{equation}}

\def\br{{\bf r}}
\def\bq{{\bf q}}

\begin{document}

\title{
Zero mode protection at particle-hole symmetry:\\
a geometric interpretation
}

\author{K. Ziegler}
\affiliation{Institut f\"ur Physik, Universit\"at Augsburg\\
D-86135 Augsburg, Germany\\
}
\date{\today}

\begin{abstract}
The properties of zero modes in particle-hole symmetric systems are analyzed
in the presence of strong random scattering by a disordered environment. The study 
is based on the calculation of the time-averaged density distribution on a 
lattice. In particular, a flat distribution is found for strong random
scattering. This result is compared with a decaying distribution for weak
random scattering by an analysis of the scattering paths. 
In the calculation we consider the 
invariant measure of the average two-particle Green's function, which is 
related to lattice-covering self-avoiding (LCSA) strings. In particular, 
strong scattering is associated with LCSA loops, 
whereas weaker scattering is associated with open LCSA strings. 
Our results are a generalization of the delocalized state observed at the band center 
of a one-dimensional tight-binding model with random hopping by Dyson in 1953.
\end{abstract}

\maketitle

\section{Introduction}

We consider the unitary evolution of a quantum system, which is characterized by the Hamiltonian $H$.
Then the time-averaged transition probability $P_{\br\br'}$ between two quantum states $|\br\rangle\to|\br'\rangle$ 
is defined as
\beq
P_{\br\br'}=\lim_{T\to\infty}\frac{1}{T}\int_0^T\Big|\langle\br'|e^{-iHt}|\br\rangle\Big|^2e^{-t/T}dt
=\frac{1}{\pi}\lim_{\epsilon\to0}\epsilon\int G_{\br\br'}(E+i\epsilon)G_{\br'\br}(E-i\epsilon)dE
\eeq
with $\epsilon=1/T$ and the Green's function $G_{\br\br'}(E+i\epsilon)=\langle\br'|(H-E-i\epsilon)^{-1}|\br\rangle$. 
In a fermionic system at low temperature only states near the Fermi surface
contribute to transport. Therefore, the main contribution to $P_{\br\br'}$ comes from the 
two-particle Green's function
\beq
\pi_{\br\br'}=\frac{1}{\pi}\lim_{\epsilon\to0}\epsilon
G_{\br\br'}(E_F+i\epsilon)G_{\br'\br}(E_F-i\epsilon)
\ ,
\label{2pgf}
\eeq
where $E_F=0$ for a particle-hole symmetric system with $H\to -H$ under particle-hole
transformation. On a lattice the state $|\br\rangle$ is a local Wannier state at the site 
$\br$ and $\pi_{\br\br'}$ describes the spatial spreading of the particle density from the source at $\br'$ 
to a site $\br$ which can be considered as a time-averaged density distribution (DD) of the quantum system. 
Expression (\ref{2pgf}) can be understood as a classical interpretation of the quantum dynamics in a disordered environment.  
It has been the foundation of transport in disordered quantum systems, based on the Kubo formalism,
and has attracted attention by researchers of different background, ranging from electronic to photonic
systems
\cite{haldane08,raghu08,cheng16}. 
For the latter the time-averaged intensity of a monochromatic electromagnetic field with frequency $\omega$,
created by a local source at $\br_0=0$, reads \cite{mishchenko06,ziegler17a,ziegler17b}
\beq
I(\br,\omega)
=\lim_{\epsilon\to 0}\epsilon G_{\br0}(\omega+i\epsilon)G_{0\br}(\omega-i\epsilon) |j_{0;1}|^2
\ .
\label{intensity000}
\eeq
$|j_{0;1}|^2$ is the intensity of the local source. 

The connection of the long-range properties of $\langle\pi_{\br\br'}\rangle_d$, where $\langle ...\rangle_d$
is the average over a random distribution of scatterers, with symmetries and spontaneous 
symmetry breaking was discussed early on by Wegner within a functional integral approach
\cite{wegner79}, who realized that this aspect is related to a nonlinear sigma model \cite{wegner80}. 
The latter can be derived from the invariant measure in a gradient expansion up to second order.
This concept can also be applied to two-dimensional particle-hole symmetric systems. 
Particle-hole symmetry is important for a large class of physical systems, ranging from superconductors
over Dirac fermions to topological materials. 
In contrast to the general approach of Wegner, the particle-hole symmetry implies a reduction of the
underlying integration space \cite{ziegler97}.

Following the standard integration procedure for disordered systems,
we must replicate the integration space of both Green's function $(H\pm i\epsilon)^{-1}$ separately, 
either using a fermion-boson pair or $n$ fermion or boson replicas. The reason for replicating both
Green's functions separately is that in general $\det(H+i\epsilon)\ne \det(H-i\epsilon)$.
On the other hand, in the particle-hole symmetric case with the particle-hole transformation 
$UH^TU^\dagger=-H$ ($U$ is a unitary transformation, $^T$ is the matrix transposition),
it is sufficient to replicate only $(H+i\epsilon)^{-1}$,
since $\det(H-i\epsilon)=\det(-H+i\epsilon)=\det(H^T+i\epsilon)=\det(H+i\epsilon)$. The first
equation follows from the fact that $H$ is assumed to be $2N\times2N$ matrix. This enables us to
choose, for instance, fermions for $(H+i\epsilon)^{-1}$ and bosons for $(H^T+i\epsilon)^{-1}$.
Then it turns out that a rotation in the fermion-bose space is a symmetry transformation
that is broken only by the $\epsilon$ term, and the above mentioned spontaneous symmetry breaking 
is found in the limit $\epsilon\to 0$. There exists a Grassmann submanifold in the integration space, 
which is associated with the spontaneously broken symmetry. This is described by an invariant measure, 
which is the Jacobian for the integration on the the associated submanifold. It depends only 
on a Grassmann field \cite{Ziegler2009,Ziegler2009a}. 

Employing the nonlinear sigma model approximation by expanding the invariant measure
up to second order in the gradient operator leads to a diffusive behavior in the case of
two-dimensional Dirac fermions with diffusion coefficient
\beq
D(E)=\lim_{\epsilon\to0}\epsilon^2\sum_\br  r_k^2\langle G_{\br0}(E+i\epsilon)G_{0\br}(E-i\epsilon)\rangle_d
\ .
\eeq
This is not surprising for weak disorder because massless two-dimensional Dirac fermions are diffusive already in the 
absence of disorder. For stronger disorder this approximation is not sufficient though, as previous
calculations have indicated \cite{1404}. In other words, the evaluation of the average two-particle 
Green's function
\beq
K_{{\bar r}{\bar r}'}\sim \langle G_{+;{\bar r}{\bar r}'}G_{-;{\bar r}'{\bar r}}\rangle_d
\ \ {\rm with} \ \ 
G_\pm=G(\mp i\epsilon)=(H\pm i\epsilon)^{-1}
\ ,
\label{2pgf00}
\eeq
must be based on the invariant measure, which neglects exponentially decaying contributions and keeps 
only the long-range properties. The physics of scattering in $K_{{\bar r}{\bar r}'}$ can also be visualized 
graphically. The individual Green's functions in Eq. (\ref{2pgf}) can be represented by Feynman path integrals
\cite{feynman,glimm,schulman}
for a particle moving from $\br'$ to $\br$ and by the complex conjugate integral for a particle moving
from $\br$ to $\br'$ (reversed path). This pair of paths must be averaged with respect to the disorder
distribution to get Eq. (\ref{2pgf00}). The averaging procedure is affected by strong interference,
since the Feynman paths are weighted by complex phase factors. As a result of the Grassmann field in 
the invariant measure, the corresponding paths of $K_{{\bar r}{\bar r}'}$  are LCSA strings. Details 
of these strings, depending on the scattering strength, are discussed in this paper.

The paper is organized as follows. In Sect. \ref{sect:av-2pgf} the average two-particle 
Green's function $K_{{\bar r}{\bar r}'}$ is represented by the invariant measure and by a Grassmann
functional integral with random phases. Then the strong random scattering asymptotics is calculated in
Sect. \ref{sect:sse1}, and in Sect. \ref{sect:geom_int} a geometric interpretation of the LCSA strings
for strong random scattering (LCSA loops) and weak random scattering (open LCSA strings) are compared.
Finally, in Sect. \ref{sect:discussion} the results of the calculation are discussed in a more general 
context. 

\section{Averaged two-particle Green's function}
\label{sect:av-2pgf}

Now we consider a spinor Hamiltonian on a lattice of $N$ sites and restrict ourselves to
a two-component spinor space. Then the Hamiltonian can be expanded in the Pauli matrix basis 
as $H=h_1\sigma_1+h_2\sigma_2+h_3\sigma_2$, where the coefficients $h_j$ are $N\times N$
matrices on the lattice.
The effective average one-particle Green's function at $E=0$ (i.e., at the symmetry point between two bands)
reads for the random $2N\times2N$ Hamiltonian $H$ in self-consistent Born approximation
\beq
\langle (H\pm 2i\epsilon)^{-1}\rangle_d\approx g_\pm
=[H_0\pm 2i(\epsilon+\eta)]^{-1}
\ , \ \ 
H_0=\langle H\rangle_d +\Sigma'
\eeq
where $\Sigma'$ is the real part of the self energy and $2\eta$ is its imaginary part.

The effective Green's function $g_\pm$ is the starting point of our study.
It has a simple physical interpretation, in which $H_0$ describes the propagation of a quantum
particle on a lattice and $\eta$ is an effective scattering rate caused by random scattering.
$\eta$ is proportional to the density of states at the Fermi level $E_F=0$ \cite{ziegler97}. 
Thus, it is directly related to the average one-particle Green's function;
i.e., it is the imaginary part of $g_-$.
The result can be used as an input to evaluate the average 
two-particle Green's function (\ref{2pgf00}) as the Grassmann functional integral \cite{berezin,negele,ziegler12}
\beq
K_{{\bar r}{\bar r}'}
=-\frac{1}{\cal N}\int_{\cal G} \varphi_{\bar r}\varphi_{{\bar r}'}' J
\ , \ \ 
{\cal N}=\int_{\cal G} J
\label{corr00}
\eeq
with the Jacobian $J$ that reads
\beq
J^{-1}=
\det({\bf 1}+\varphi \varphi'-\varphi h\varphi'h^\dagger)
\ ,\ \
h:={\bf 1}-4i\eta g_+
\ ,\ \
h^\dagger:={\bf 1}+4i\eta g_-
\ .
\label{jacobian00}
\eeq
Using the definition of $h$ the inverse Jacobian also reads
\beq
J^{-1}=\det[{\bf 1}+4i\eta(\varphi g_+\varphi'-\varphi\varphi' g_-)-16\eta^2\varphi g_+\varphi'g_-]
\ .
\label{jacobian00a}
\eeq
It was shown in Ref. \cite{1404} that the inverse determinant $J$ can be expressed with the help of
a random phase $\alpha_{\br j}$ (i.e., with a complex bosonic field) 
by the replacement $H_0\to{\bf H}$ with 
\beq
{\bf H}_{\br j,\br'j}=e^{i\alpha_{\br j}}H_{0;\br j,\br'j'}e^{-i\alpha_{\br'j'}}
\ .
\eeq
Then the averaged two-particle Green's function becomes a phase averaged Gaussian Grassmann integral
\beq
K_{{\bar \br}{\bar \br}'}
=-\frac{1}{\cal N}\Big\langle\int_{\cal G} \varphi_{{\bar r}'}'\varphi_{\bar r}
\exp\left(\sum_{\br,\br'}\varphi_\br C_{\br\br'}\varphi_{\br'}'\right) 
\Big\rangle_\alpha
=\frac{\langle adj_{{\bar \br}{\bar \br}'}C\rangle_\alpha}{{\cal N}}
\ ,\ \
\langle ...\rangle_\alpha=\frac{1}{2\pi}\int_0^{2\pi}...\prod_{\br,j}d\alpha_{\br j}
\label{corr1a}
\eeq
with the adjugate matrix $adj_{{\bar \br}{\bar \br}'} C$ and with the normalization factor
\beq
{\cal N}=\Big\langle \int_{\cal G} \exp\left(\sum_{\br,\br'}\varphi_\br C_{\br\br'}\varphi_{\br'}'\right) 
\Big\rangle_\alpha
=\langle\det C\rangle_\alpha
\ .
\label{normal1}
\eeq
The $N\times N$ random phase matrix $C$, whose elements are
\beq
C_{\br\br'}
=4i\eta\sum_{j,j'}\left({\bf g}_{\br j,\br' j'}-\delta_{\br\br'}\sum_{\br''}{\bf g}^\dagger_{\br j,\br'' j'}\right)
-16\eta^2\sum_{j,j'}{\bf g}_{\br j,\br'j'}\sum_{\br'',j''}{\bf g}^\dagger_{\br'j',\br''j''}
\eeq
with 
${\bf g}=[{\bf H}+ 2i{\bar\eta}]^{-1}$, ${\bar\eta}=\eta+\epsilon$, determines the properties of the DD.
With $\kappa_\br= 16\eta\sum_{\br'}\sum_{j,j'}({\bf g}{\bf g}^\dagger)_{\br j,\br' j'}$
and with the relation ${\bf g}-{\bf g}^\dagger=-4i{\bar\eta}{\bf g}{\bf g}^\dagger$ we get
a more compact version of $C$ as
\beq
C_{\br\br'}
=\left(\epsilon\kappa_\br-\sum_{\br''}\Delta_{\br\br''}\right)\delta_{\br\br'}+\Delta_{\br\br'}
\ ,\ \ \
\Delta_{\br\br'}=4i\eta\sum_{j,j'}{\bf g}_{\br j,\br'j'}\left(1
+4i\eta\sum_{\br'',j''}{\bf g}^\dagger_{\br'j',\br''j''}\right)
\ .
\label{C_corr}
\eeq
The form of the matrix $C_{\br\br'}$ implies 
$
\sum_{\br'}C_{\br\br'}=\epsilon\kappa_\br
$,
such that for $\epsilon=0$ there is a zero eigenvalue with a constant eigenvector, regardless of the 
specific realization of the random ${\bf g}$. Thus, $\det C$ always vanishes
in the limit $\epsilon\to 0$. The very existence of this zero mode 
is essential for the behavior of the two-particle Green's function. What remains to be discussed is 
the related propagation for different wavevectors, which is the subject of this paper.
For the strong-scattering asymptotics it is easier to start directly from Eq. (\ref{jacobian00a}),
while for a general discussion in Sect. \ref{sect:geom_int} the random approach of
Eqs. (\ref{corr1a}), (\ref{normal1}) is more convenient.

\section{Strong-scattering asymptotics}
\label{sect:sse1}

Starting from the Jacobian $J$, defined in Eq. (\ref{jacobian00a}) on a lattice with $N$ sites,
we expand 
\[
4i\eta(\varphi g_+\varphi'-\varphi\varphi' g_-)-16\eta^2\varphi g_+\varphi'g_-
\]
with $\eta/{\bar\eta}\sim1$ as
\beq
4\frac{\epsilon}{{\bar\eta}}\varphi\varphi'-i\frac{1}{\bar\eta}(\varphi H_0\varphi'-\varphi\varphi'H_0)
+\frac{1}{2{\bar\eta}^2}(\varphi H_0^2\varphi'+\varphi\varphi'H_0^2)-\frac{1}{{\bar\eta}^2}\varphi H_0\varphi'H_0
+O({\bar\eta}^{-3})
\ ,
\label{expan01}
\eeq
and rescale the Grassmann field $\varphi\varphi'/{\bar\eta}\to\varphi\varphi'$ such that we obtain
\beq
-i[\varphi H_0\varphi'-\varphi\varphi'(H_0-4i\epsilon)]
+\frac{1}{2{\bar\eta}}(\varphi H_0^2\varphi'+\varphi\varphi'H_0^2)-\frac{1}{{\bar\eta}}\varphi H_0\varphi'H_0
+O({\bar\eta}^{-2})
\ . 
\eeq
Then we get in leading order of $1/{\bar\eta}$ 
\[
-\log J=
-\sum_{l=1}^N\frac{(-1)^l}{l}\sum_{\{\br_k,j_k\}, 1\le k\le l} 
A^{(l)}_{\br_1,j_1,...,\br_{l+1}, j_{l+1}}\Big|_{\br_{l+1}=\br_1,j_{l+1}=j_1}
\ ,
\]
where
\[
A^{(l)}_{\br_1,j_1,...,\br_{l+1} j_{l+1}}
=i^{l}\Big[
\varphi_{\br_1}\varphi_{\br_1}'H_{0;\br_1 j_1,\br_2 j_2}'\cdots\varphi_{\br_{l}j_{l}}\varphi_{\br_{l}j_{l}}'
H_{0;\br_{l} j_{l},\br_{l+1} j_{l+1}}'
\]
\[
-\varphi_{\br_{l+1}}\varphi_{\br_1}'H_{0;\br_1 j_1,\br_2 j_2}\cdots\varphi_{\br_{l}j_{l}}\varphi_{\br_{l}j_{l}}'
H_{0;\br_{l} j_{l},\br_{l+1} j_{l+1}}\Big]
\]
with $H_{0;\br j,\br' j'}'=H_{0;\br j,\br' j'}-4i\epsilon\delta_{\br\br'}\delta_{jj'}$.
For $\br_{l+1}=\br_1$ and $j_{l+1}=j_1$ (i.e., for a loop) we obtain
\beq
A^{(l)}_{\br_1,j_1,...,\br_1 j_1}
=\cases{
4\epsilon\varphi_{\br_1}\varphi_{\br_1}' & $l=1$  \cr
0 & $2\le l\le N$ \cr
}
\ .
\label{leading_sum}
\eeq
Thus, the leading order expression results in the simple relation
\beq
Tr\left[\log\left({\bf 1}+\varphi \varphi'-\varphi h\varphi'h^\dagger\right)\right]
=8\epsilon\sum_\br\varphi_{\br}\varphi_{\br}'+O({\bar\eta}^{-1})
\ ,
\eeq
for which the Grassmann integration gives
\[
\int_{\cal G}\exp\left\{-Tr\left[\log\left({\bf 1}+\varphi \varphi'-\varphi h\varphi'h^\dagger\right)\right]\right\}
\sim (-8\epsilon)^N
\ .
\]
Next we consider a perturbation of order $1/{\bar\eta}$. A lengthy but straightforward calculation 
(App. \ref{app:expansion}) gives for $\epsilon\sim0$
\beq
\int_{\cal G}\det\left({\bf 1}+\varphi \varphi'-\varphi h\varphi'h^\dagger\right)^{-1}
\sim (-8\epsilon)^N
+\frac{8\epsilon}{\bar\eta}i^{N-2}\int_{\cal G} Tr\left[(\varphi\varphi'H_0)^N\right]
+O({\bar\eta}^{-2})
\ ,
\label{det11}
\eeq
which represents an LCSA loop, and with Eq. (\ref{sum13}) we obtain with the spinor trace $Tr_2$
\beq
\int_{\cal G}\varphi_{\bar\br}\varphi_{{\bar\br}'}'
\det\left({\bf 1}+\varphi \varphi'-\varphi h\varphi'h^\dagger\right)^{-1}
\sim -\frac{i^{N-2}}{\bar\eta}\sum_{k=0}^{N-2}\int_{\cal G} 
Tr_2
\left[(\varphi\varphi'H_0)^{N-1-k}_{{\bar\br}{\bar\br}'}(\varphi\varphi'H_0)^{k+1}_{{\bar\br}'{\bar\br}}
\right]+O({\bar\eta}^{-2})
\ ,
\label{corr11}
\eeq
which represents an LCSA loop with fixed positions $\bar\br$ and ${\bar\br}'$
with ${\bar\br}'\ne{\bar\br}$. For the diagonal case ${\bar\br}'={\bar\br}$ we get
\beq
\int_{\cal G}\varphi_{\bar\br}\varphi_{{\bar\br}}'
\det\left({\bf 1}+\varphi \varphi'-\varphi h\varphi'h^\dagger\right)^{-1}
\sim (-8\epsilon)^{N-1} -\frac{i^{N-2}}{{\bar\eta}}\int_{\cal G} Tr\left[(\varphi\varphi'H_0)^N\right]
+O({\bar\eta}^{-2})
\ .
\eeq
For ${\bar\br}'\ne{\bar\br}$
the summation over the coordinates $\{\br_k\}$ ($k=1,...,N$) is actually a (partial) permutation of the lattice
sites due to the Grassmann factors $\varphi\varphi'$. This implies that the $k$ summation of Eq. (\ref{corr11})
is identical to the integral in Eq. (\ref{det11}), except for a factor $1/N$: 
\beq
D:=i^{N-2}\sum_{k=0}^{N-2}\int_{\cal G} Tr_2
\left[(\varphi\varphi'H_0)^{N-1-k}_{{\bar\br}{\bar\br}'}(\varphi\varphi'H_0)^{k+1}_{{\bar\br}'{\bar\br}}
\right]
=\frac{i^{N-2}}{N}\int_{\cal G} Tr\left[(\varphi\varphi'H_0)^N\right]
\ .
\label{D00}
\eeq
This follows from the fact that (i) we have for a translational invariant $H_0$
\[
Tr\left[(\varphi\varphi'H_0)^N\right]=NTr_2[(\varphi\varphi'H_0)^N_{{\bar\br}{\bar\br}}]
\]
and (ii) the sum over permutations of lattice sites reads for ${\bar\br}\ne{\bar\br}'$
\beq
(\varphi\varphi'H_0)^N_{{\bar\br}{\bar\br}}
=\sum_{m=1}^{N-1}(\varphi\varphi'H_0)^{N-m}_{{\bar\br}{\bar\br}'}(\varphi\varphi'H_0)^{m}_{{\bar\br}'{\bar\br}}
\ ,
\eeq
which gives Eq. (\ref{D00}). Eventually we obtain
\beq
K_{\br\br'}=\frac{{\bar\eta}}{8\epsilon}\frac{\delta_{\br\br'}(-8\epsilon)^{N-1}-{\bar\eta}^{-1}D+O({\bar\eta}^{-2})}
{(-8\epsilon)^{N-1}-N{\bar\eta}^{-1}D+O({\bar\eta}^{-2})}
\ ,
\label{K_final}
\eeq
where the prefactor ${\bar\eta}$ is the result of the rescaling of the Grassmann field.
In the limit ${\bar\eta}\to\infty$ 
we get strong Anderson localization with vanishing localization length
\beq
K_{\br\br'}\sim\frac{{\bar\eta}}{8\epsilon}\delta_{\br\br'}
\ ,
\label{AL}
\eeq
whereas for a large but finite ${\bar\eta}$ with ${\bar\eta}\ll \epsilon (E_t/\epsilon)^N$
($E_t$ is the tunneling energy of the Hamiltonian $H_0$) we get
\beq
K_{\br\br'}\sim\frac{{\bar\eta}}{8N\epsilon}
\ .
\label{fin_res}
\eeq
This result describes a uniform DD over the entire lattice.

\begin{figure}[t]
\begin{center}
\includegraphics[width=8cm,height=3.5cm]{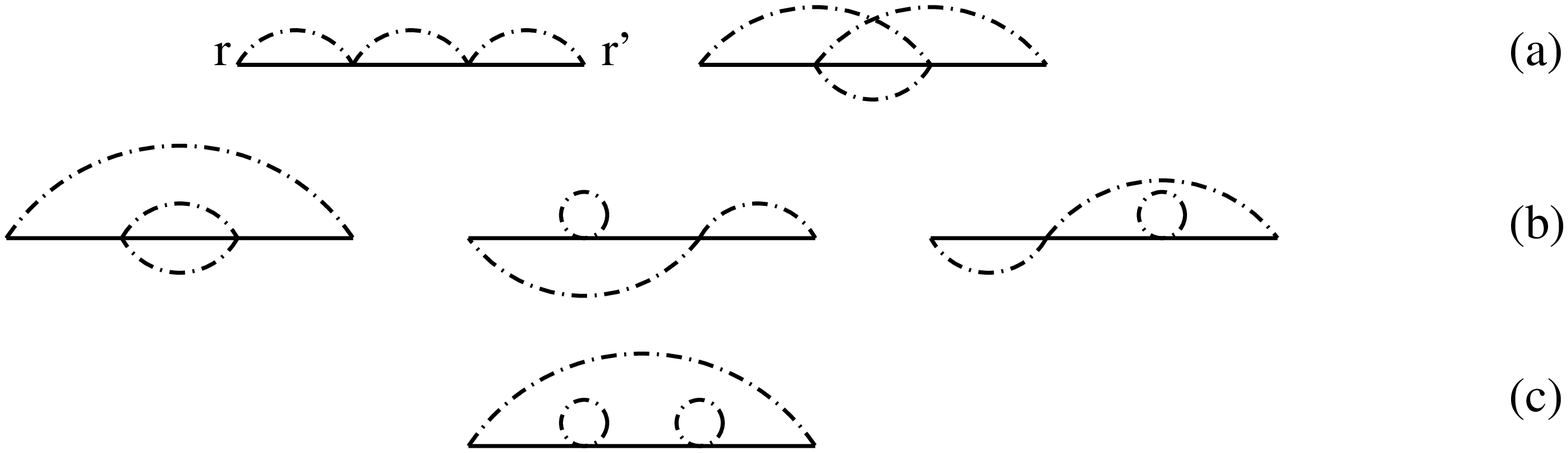}
\caption{
Graphs of $K_{\br\br'}$ with open LCSA strings after phase averaging for 4 sites. Dashed lines represent 
$g_-$, full lines $g_+$. There is one loop in (a), two loops in (b), and three loops in (c).   
}
\label{fig:graphs}
\end{center}
\end{figure}

Although $D$ drops out in the first order perturbation,
for a deeper insight into the physical origin of the result (\ref{fin_res}) we must calculate $D$.
The spinor trace $Tr_2$ can be rewritten in terms of phase factors or Ising spins: For the product
of $2\times2$ matrices $H_{0;\br_k,\br_{k+1}}$ we replace them by the scalars
${\tilde A}_{\br_k\br_{k+1}}=\sum_{j,j'}S_{\br j}H_{0;\br_k j,\br' j'}S_{\br_{k+1}j'}^*$, 
where $S_{\br j}$ is either a phase factor $\exp(i\alpha_{\br j})$ ($0\le\alpha_{\br j}<2\pi$) \cite{1404}
or an Ising spin ($S_{\br j}=\pm 1$). For simplicity we choose the Ising spin here. Then the trace reads
\[
Tr_2(H_{0;\br_1\br_2}\cdots H_{0;\br_{n}\br_1})
=2^{-2n}\sum_{\{ S_{\br_k j}=\pm1\}}({\tilde A}_{\br_1\br_2}\cdots {\tilde A}_{\br_{n}\br_1})
\ .
\]
For the example $H_{0}=h_1\sigma_1+h_2\sigma_2$ we obtain
\[
{\tilde A}_{\br\br'}=S_{\br 1}S_{\br' 2}\left( a_{\br\br'}+T_\br b_{\br\br'}T_{\br'}\right)
\ ,\ \ 
T_\br=S_{\br 1}S_{\br 2}
\ , \ \ 
a=h_1-ih_2 ,\ b=h_1+ih_2
\]
and
\[
2^{-2n}\sum_{\{ S_{\br_k j}=\pm1\}}{\tilde A}_{\br_1\br_2}\cdots {\tilde A}_{\br_{n}\br_1}
=a_{\br_1\br_2}\cdots a_{\br_n\br_1}+b_{\br_1\br_2}\cdots b_{\br_n\br_1}
\]
such that we rewrite
\beq
D
=\frac{1}{N}{\sum_{\br_1,...,\br_N}}'
(a_{\br_1\br_2}\cdots a_{\br_N\br_1}+b_{\br_1\br_2}\cdots b_{\br_N\br_1})
\ ,
\label{det12}
\eeq
where ${\sum_{\br_1,...,\br_N}}'$ is the restricted sum with $\br_j\ne\br_k$ for $j\ne k$.
Thus, $D$ is represented by the sum of two types of LCSA loops,
where the ``propagators'' are $a=h_1-ih_2$ and $b=h_1+ih_2$, respectively.
\begin{figure}[t]
\begin{center}
\includegraphics[width=11cm,height=4cm]{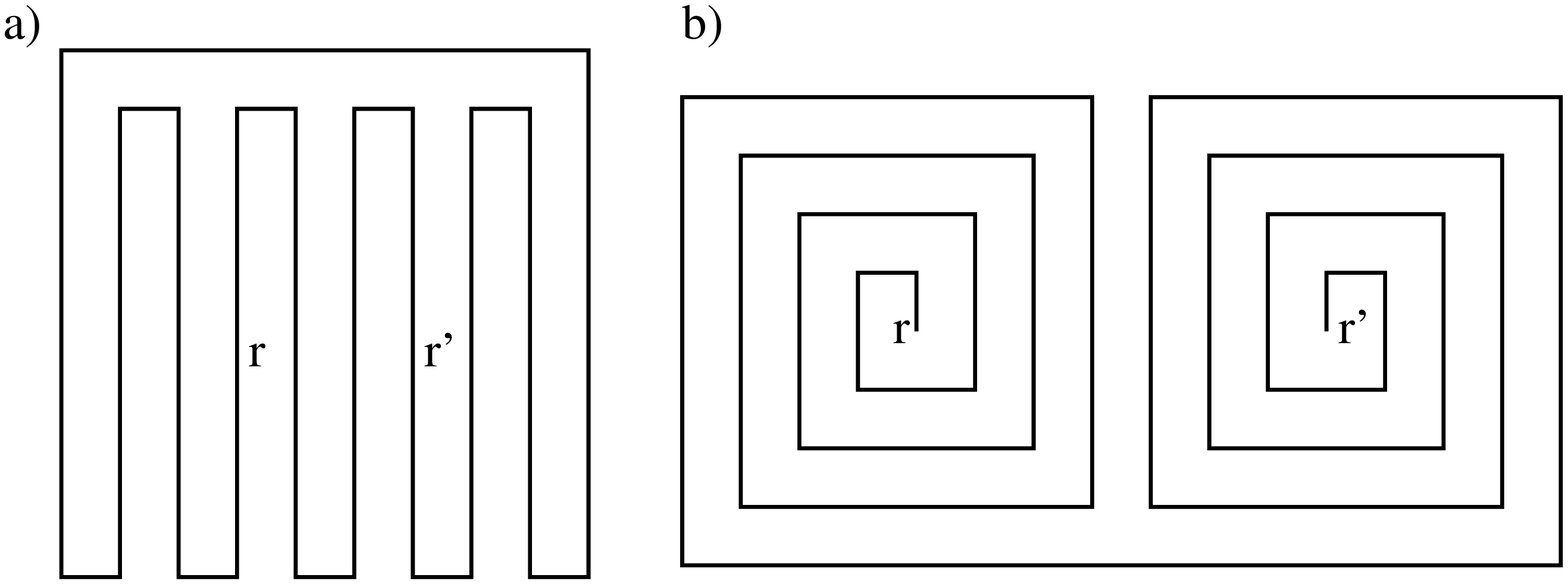}
\caption{
Examples for LCSA strings: 
a) a loop from the strong scattering limit of Eq. (\ref{corr11}) and b) an open string 
from the truncated expansion for $adj_{\br\br'}{\bar C}$ of Eq. (\ref{string_off})
that connects the sites $\br'$ and $\br$. 
}
\label{fig:strings}
\end{center}
\end{figure}

\section{Geometric interpretation}
\label{sect:geom_int}

We have found in Sect. \ref{sect:sse1} that the asymptotic behavior of the DD is governed 
at strong scattering by an LCSA loop (cf. Fig. \ref{fig:strings}a). 
On the other hand, for moderate scattering 
the DD is related to LCSA string with 4-vertices \cite{1404}. An example is given for a lattice with
four sites in Fig. \ref{fig:graphs}. 
The 4-vertex graphs resemble the thermal statistics of the two-dimensional Ising model at the critical point
\cite{mccoy,itzykson}. In the following we shall study and compare different contributions to the DD
which are characterized according to their geometric structure as LCSA strings. While the actual form 
and values of the propagators are not relevant, the shape of the LCSA strings is essential. In
particular, a single loop vs. an open string that is entangled with other strings belong to different
classes (cf. Fig. \ref{fig:strings}a and \ref{fig:strings}b).

In general, an open string from $\br$ to $\br'$ on a lattice with $n+2$ sites (cf. Eq. (\ref{string11})), 
\[
\Delta_{\br\br_1}\Delta_{\br_1\br_2}\cdots \Delta_{\br_n\br'}
[-\varphi_{\br'}'\varphi_\br+\varphi_\br'\varphi_\br]
\varphi_{\br_1}'\varphi_{\br_1} \varphi_{\br_2}' \cdots \varphi_{\br_n}
\]
cannot cover the entire lattice with Grassmann variables.
This means that we need additional contributions with Grassmann variables. To find those we
must distinguish the Grassmann integral of ${\cal N}$ in Eq. (\ref{normal1}) and the corresponding
Grassmann integral for $K_{{\bar\br}{\bar\br}'}$ in Eq. (\ref{corr1a}). Beginning with ${\cal N}$,
we can only use the diagonal term from the square brackets of Eq. (\ref{string11})
\[
\Delta_{\br\br_1}\Delta_{\br_1\br_2}\cdots \Delta_{\br_n\br'}
\varphi_{\br}'\varphi_\br
\varphi_{\br_1}'\varphi_{\br_1} \varphi_{\br_2}' \cdots \varphi_{\br_n}
\ ,
\]
set $\br_n=\br$ and borrow a factor $\epsilon\kappa_{\br'}\varphi_{\br'}\varphi_{\br'}'$ from the 
expansion of $\exp(-\epsilon\sum_{\br}\kappa_\br\varphi_\br\varphi_\br')$. This implies that
${\cal N}$ always vanishes with $\epsilon\to0$, as already mentioned at the end Sect. \ref{sect:av-2pgf}.
Geometrically this is an LCSA loop as visualized in Fig. \ref{fig:strings}a.
Moreover, averaging over the random phases of the LCSA string generates additional bridges between lattice sites
due to the term $\sum_{\br'',j''}{\bf g}^\dagger_{\br'j',\br''j''}$ in (\ref{C_corr}),
such that we eventually get a LCSA graph with 4-vertices at each site \cite{1404}.
These bridges do not occur in the strong scattering limit though, as discussed in the previous section.
Thus, the $1/{\bar\eta}$-expansion is graphically an expansion in terms of 4-vertices. 

In the case of the integral (\ref{corr1a}) we must take the off-diagonal term from the square
brackets in Eq. (\ref{string11}),
\[
-\Delta_{\br\br_1}\Delta_{\br_1\br_2}\cdots \Delta_{\br_n\br'}
\varphi_{\br'}'\varphi_\br
\varphi_{\br_1}'\varphi_{\br_1} \varphi_{\br_2}' \cdots \varphi_{\br_n}
\]
and set $\br={\bar\br}'$, $\br'={\bar\br}$. Thus the extra factor $\varphi_{{\bar r}'}'\varphi_{\bar r}$
in Eq. (\ref{corr1a}) completes the lattice covering by Grassmann variables. 
At fixed ${\bar\br}$, ${\bar\br}'$ we get the expression
\beq
(\varphi_{{\bar\br}'}\Delta_{\br\br_1}\varphi_{\br_1}')(\varphi_{\br_1}
\Delta_{\br_1\br_2}\varphi_{\br_2}')\cdots (\varphi_{\br_n}\Delta_{\br_n\br'}
\varphi_{\bar\br}')
\label{string_off}
\eeq
which consists of factors of $\varphi_{\br_k}\Delta_{\br_k\br_{k+1}}\varphi_{\br_{k+1}}'$
along a string connecting ${\bar\br}'$ and ${\bar\br}$, as depicted in Fig. \ref{fig:strings}b.
This string must still be averaged with respect to the random phases. Again, averaging creates bridges 
between the lattice sites, which results in a graph with 4-vertices \cite{1404}.
A simple example is depicted in Fig. \ref{fig:graphs} which has 4-vertices except for the two
external sites $\br$ and $\br'$ with 2-vertices. The full line is the open LCSA string and the dashed lines
are the bridges created by phase averaging. 
To compare this with the strong scattering asymptotic of Sect. \ref{sect:sse1}, we
can create a loop from an open string with bridges through phase averaging.
As a result we get the loop of Fig. \ref{fig:strings}a. An example with four lattice sites 
is the graph in Fig. \ref{fig:graphs}(c): after a $1/{\bar\eta}$ expansion we obtain
a loop of the type in Fig. \ref{fig:strings}a, since both $g_-$ and $g_+$ contribute an $H_0$ in 
the leading terms of the expansion. 

\begin{figure}[t]
\begin{center}
\includegraphics[width=7cm,height=3.5cm]{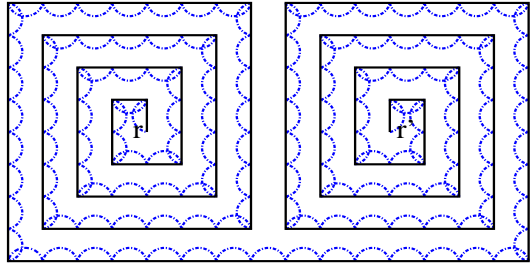}
\caption{
Example for a loop that consists of an open LCSA string connecting $\br'$ and $\br$ which is connected with a 
sequence of blue dashed bridges. The bridges are created by phase averaging. The links along the LCSA string 
consist of $g_+$ propagators and the bridges consist of $g_-$ propagators.
}
\label{fig:strings2}
\end{center}
\end{figure}

Phase averaging creates different configurations of bridges between lattice sites. A simple case is
one in which the bridges connect neighboring sites along the string, as visualized in left graph of 
Fig. \ref{fig:graphs}a and in Fig. \ref{fig:strings2}. 
(This usually does not mean nearest neighbors on the lattice.)  This case is formally associated with the   
replacement $\Delta_{\br_k\br_{k+1}}$ by $\langle\Delta_{\br_k\br_{k+1}}\rangle_\alpha$ in 
(\ref{string_off}). The same result is obtained from the expression (\ref{jacobian00a})
by expanding $-\log J$ up to first order (also known as the nonlinear sigma model approximation of 
the invariant measure):
\[
-\log J\approx 
16\eta{\bar\eta} \sum_\br Tr_2(g_+g_-)_{\br\br}\varphi_\br\varphi_\br'
-16\eta^2\sum_{\br,\br'}Tr_2(g_{+;\br\br'}g_{-;\br'\br})\varphi_\br \varphi_{\br'}'
\]
which contains the quadratic form $\sum_{\br,\br'}\varphi_\br {\bar C}_{\br\br'}\varphi_{\br'}'$
with
\[
{\bar C}_{\br\br'}=\langle C_{\br\br'}\rangle
=16\eta\epsilon Tr_2(g_+g_-)_{\br\br}\delta_{\br\br'}
+16\eta^2\left[
\sum_{\br''}Tr_2(g_{+;\br\br''}g_{-;\br''\br})\delta_{\br\br'}
- Tr_2(g_{+;\br\br'}g_{-;\br'\br})\right]
\ .
\]
For this example we obtain an approximation of the DD (\ref{corr1a}) as 
\beq
K_{{\bar\br}{\bar\br}'}\approx {\bar C}^{-1}_{{\bar\br}{\bar\br}'}
\ .
\label{decay}
\eeq
Thus, the DD decays according to a power law: 
After a Fourier transformation ${\bar C}_{\br-\br'}\to {\tilde C}_\bq$
we get
$
{\tilde C}_\bq
\sim D q^2
$
($
q\sim0
$)
and
\beq
K_{{\bar\br}{\bar\br}'}\to {\tilde K}_\bq=\frac{1}{{\bar\kappa}\epsilon +D q^2+O(q^3)}
\label{nlsm_result}
\eeq
with
\[
{\bar\kappa}=16\eta Tr_2(H_0^2+4{\bar\eta}^2)^{-1}_{\br\br}
\ ,\ \ 
D=16\eta^2\sum_{\br}r^2
Tr_2(g_{+;\br-\br'}g_{-;\br'-\br})
\ .
\]

Another example of the random phase representation of the Jacobian in Eqs. (\ref{corr1a}), (\ref{normal1})
uses the mean-phase approximation \cite{1404,ziegler17b}. Since we do not average this leads to a decaying 
DD as given in Eq. (\ref{decay}) where ${\bar C}_{\br\br'}$ is replaced by $C_{\br\br'}$
of Eq. (\ref{C_corr}) with fixed uniform phases: $e^{i\alpha_j}$ ($j=1,2$), where the two phase parameters 
$\alpha_{1,2}$ are determined by a variational principle \cite{1404}:
\[
H_{0;\br j,\br' j'}\to {\cal H}_{0;\br j,\br' j'}=H_{0;\br j,\br' j'}e^{i(\alpha_j-\alpha_{j'})}
\ .
\]
The phase factors can also be understood as a rotation when we use a representation in terms of 
Pauli matrices $H_0=h_1\sigma_1+h_2\sigma_2 +h_3\sigma_3$:
\[
{\cal H}_0=(h_1c+h_2s)\sigma_1+(h_2c-h_1s)\sigma_2 +h_3\sigma_3
\ ,\ \ \
c=\cos(\alpha_1-\alpha_2)
\ ,\ \
s=\sin(\alpha_1-\alpha_2)
\ 
\]
describes a rotation of $H_0$ around the $\sigma_3$ axis by the angle $\alpha_1-\alpha_2$.
Since this angle is fixed by a variational principle, the discrete isotropy of the underlying lattice
would determine it at discrete values.
For $\Delta_{\br\br'}$ in Eq. (\ref{C_corr}) this implies
\[
\Delta_{\br\br'}=-8i\eta\left[(H_0^2+4{\bar\eta}^2)^{-1}_{jj}(h_1c+h_2s)\right]_{\br\br'}
-16\eta{\bar\eta}(H_0^2+4{\bar\eta}^2)^{-1}_{\br j,\br'j}
\]
In the case of two-dimensional Dirac fermions with $h_1=q_1$, $h_2=q_2$ and $h_3=m$ ($q_j$ are the components 
of the wavevector, $m$ is the Dirac mass) we obtain the result of Ref. \cite{1404}:
\beq
{\tilde K}_\bq =\frac{\tau}{4\epsilon-i{\bf s}\cdot\bq +2q^2/\tau+O(q^3)}
\ , \ \
\tau=\frac{m^2+4{\bar\eta}^2}{8\eta}
\ ,\ \ 
{\bf s}=(c,s)
\ .
\label{mpa_result}
\eeq
The term linear in the wavevector describes a ray mode along the unit vector ${\bf s}$, whereas the
quadratic term describes the diffusion of the ray mode. For strong scattering $\tau$ is large, which
favors linear propagation and suppresses the diffusive behavior. Then the expression ${\tilde K}_\bq$ 
has a pole at ${\bf s}\cdot\bq=-4i\epsilon$ which is related to the result
Eq. (\ref{fin_res}) in the direction ${\bf s}$. An additional summation (or integration) of ${\bf s}$
over all directions fixed by the variational principle approximates the constant DD.

\section{Discussion}
\label{sect:discussion}

Returning to the Feynman paths mentioned in the Introduction, we have found that the
contribution of $G_\pm$ in Eq. (\ref{2pgf00}) is replaced after averaging with respect to random
scattering by 
an open LCSA string with propagator $g_+$ from the Grassmann field that is entangled with $g_-$ strings (bridges)
through 4-vertices. Typical examples are depicted in Figs. \ref{fig:graphs}, \ref{fig:strings2}.
Contributions of the $1/{\bar\eta}$ expansion to $\langle adj_{{\bar\br}{\bar\br}'} C\rangle_\alpha$
in Eq. (\ref{K_final}) are discussed in Sect. \ref{sect:geom_int}. It is essential to all contributions that there
is a zero mode for any realization of the random phase. 

Terms of order $1/{\bar\eta}$ of the $1/{\bar\eta}$ expansion contribute equally with $H_0$ factors from $g_+$ and $g_-$  
(except for some minus signs), such that we obtain an LCSA loop (Fig. \ref{fig:strings}a). This implies the uniform DD
of Eq. (\ref{fin_res}). Another additive term to the numerator of expression (\ref{K_final}) is the partial sum 
of Fig. \ref{fig:strings2} that decays spatially according to a power law in Eq. (\ref{nlsm_result}). In this
case open LCSA strings must connect the sites ${\bar\br}$ and ${\bar\br}'$, which is the reason that $K_{{\bar\br}{\bar\br}'}$ 
decays in space unlike the constant DD in Eq. (\ref{fin_res}). And finally,
an alternative approximation for $\langle adj_{{\bar\br}{\bar\br}'} C\rangle_\alpha$ is a uniform realization of the 
random phases. It gives ray modes in the direction of the phase difference $\alpha_1-\alpha_2$ (cf. Eq. (\ref{mpa_result})). 

The main results of Sects. \ref{sect:sse1}, \ref{sect:geom_int} can be summarized as the ``absence of Anderson localization'' 
in the presence of particle-hole symmetry. The two fundamental geometric structures are visualized in Fig. \ref{fig:strings},
where Fig. \ref{fig:strings}a gives a uniform contribution to the numerator of Eq. (\ref{fin_res}) and Fig. \ref{fig:strings}b
gives a spatially decaying contribution. 
To obtain these results we must 
average over a time period $T$ ($=1/\epsilon$) which is much longer than the tunneling time $\hbar/E_t$ 
of the Hamiltonian. 

Delocalized behavior was observed for one-dimensional systems some time ago, where an extended state 
was found at the band center of a one-dimensional tight-binding model with random hopping 
\cite{dyson53,theodorou76,eggarter78}. There was also numerical evidence for an extended state at the band center in 
two-dimensional lattice models \cite{soukoulis82,inui94,CC,chalker01}.

The absence of Anderson localization is only valid {\it at} the particle-symmetry, which is just
a point in the spectrum (e.g., at the Dirac node). We suspect that away from this symmetry point there
is a finite localization length, which increases continuously as we approach the symmetry point. This 
behavior would indicate that in a vicinity of particle-hole symmetric point the localization length is 
very large. The latter could be a problem in (numerical or real) experiments to detect
the Anderson localization length. On the other hand, it could be useful for applications to avoid 
Anderson localization in disordered photonic metamaterials by tayloring the bandstructure with particle-hole 
symmetry \cite{kang18}.

\section*{ACKNOWLEDGMENT}

This work was supported by a grant of the Julian Schwinger Foundation.

\appendix

\section{Open strings 
}
\label{app:strings}

The exponential expression in Eqs. (\ref{corr1a}), (\ref{normal1}) can be rewritten with the 
help of Eq. (\ref{C_corr}) as
\[
\exp\left(\sum_{\br,\br'}\varphi_\br C_{\br\br'}\varphi_{\br'}'\right) 
=\exp(\epsilon\sum_{\br}\kappa_\br\varphi_\br\varphi_\br')
\exp(\sum_{\br,\br'}[\sum_{\br''}\Delta_{\br\br''}\delta_{\br\br'}-\Delta_{\br\br'}]
\varphi_\br\varphi_{\br'}')
\]
with
\[
\kappa_\br=16\eta Tr_2({\bf g}{\bf g}^\dagger)_{\br\br}
\ ,\ \ \
\Delta_{\br\br'}=16\eta^2 Tr_2({\bf g}_{\br\br'}{\bf g}^\dagger_{\br'\br})
\ .
\]
The expansion of the second exponential function leads to two types of open strings on the
lattice: 
type (i) consists of products of 
$\varphi_\br\varphi_\br'\sum_{\br''}\Delta_{\br\br''}$
links and type (ii) consists of products of $\varphi_\br\Delta_{\br\br'}\varphi_{\br'}'$ links.
Using the relation
\[
\exp\left[\sum_{\br,\br'}(\sum_{\br''}\Delta_{\br\br''}\delta_{\br\br'}-\Delta_{\br\br'})\varphi_\br\varphi_{\br'}'\right]
=\exp\left[\sum_{\br,\br'}\varphi_\br\Delta_{\br\br'}(\varphi_\br'-\varphi_{\br'}')\right]
=\prod_{(\br,\br')}\left[1-\varphi_\br \Delta_{\br\br'}(\varphi_{\br'}'-\varphi_\br')\right]
\]
and assuming that a string has $n+1$ links, we get
\[
[\varphi_\br \Delta_{\br\br_1}(\varphi_{\br_1}'-\varphi_\br')]
[\varphi_{\br_1} \Delta_{\br_1\br_2}(\varphi_{\br_2}'-\varphi_{\br_1}')] \cdots
[\varphi_{\br_n} \Delta_{\br_n\br'}(\varphi_{\br'}'-\varphi_{\br_n}')]
\]
\[
=\varphi_\br \Delta_{\br\br_1}\varphi_{\br_1}'
\varphi_{\br_1} \Delta_{\br_1\br_2}\varphi_{\br_2}' \cdots
\varphi_{\br_n} \Delta_{\br_n\br'}\varphi_{\br'}'
+\varphi_\br \Delta_{\br\br_1}(-\varphi_\br')
\varphi_{\br_1} \Delta_{\br_1\br_2}(-\varphi_{\br_1}') \cdots
\varphi_{\br_n} \Delta_{\br_n\br'}(-\varphi_{\br_n}')
\]
\[
=\Delta_{\br\br_1}\Delta_{\br_1\br_2}\cdots \Delta_{\br_n\br'}
[\varphi_\br \varphi_{\br_1}'\varphi_{\br_1} \varphi_{\br_2}' \cdots
\varphi_{\br_n} \varphi_{\br'}'
+\varphi_\br (-\varphi_\br')
\varphi_{\br_1} (-\varphi_{\br_1}') \cdots
\varphi_{\br_n}(-\varphi_{\br_n}')]
\]
\[
=\Delta_{\br\br_1}\Delta_{\br_1\br_2}\cdots \Delta_{\br_n\br'}
[\varphi_\br \varphi_{\br_1}'\varphi_{\br_1} \varphi_{\br_2}' \cdots
\varphi_{\br_n} \varphi_{\br'}'
+\varphi_\br'\varphi_\br
\varphi_{\br_1}'\varphi_{\br_1} \cdots
\varphi_{\br_n}'\varphi_{\br_n}]
\]
\[
=\Delta_{\br\br_1}\Delta_{\br_1\br_2}\cdots \Delta_{\br_n\br'}
[\varphi_\br (\varphi_{\br_1}'\varphi_{\br_1} \varphi_{\br_2}' \cdots
\varphi_{\br_n})\varphi_{\br'}'
+\varphi_\br'\varphi_\br
(\varphi_{\br_1}'\varphi_{\br_1} \cdots
\varphi_{\br_n}'\varphi_{\br_n})]
\]
\beq
=\Delta_{\br\br_1}\Delta_{\br_1\br_2}\cdots \Delta_{\br_n\br'}
[-\varphi_{\br'}'\varphi_\br+\varphi_\br'\varphi_\br]
\varphi_{\br_1}'\varphi_{\br_1} \varphi_{\br_2}' \cdots \varphi_{\br_n}
\ .
\label{string11}
\eeq
This expression vanishes for $\br'=\br$ (i.e., for a loop).

\section{Strong-scattering Expansion}
\label{app:expansion}

From the expression (\ref{jacobian00a})
\[
Tr\log\left[{\bf 1}+4i\eta(\varphi g_+\varphi'-\varphi\varphi' g_-)-16\eta^2\varphi g_+\varphi'g_-\right]
\]
we get with (\ref{expan01}) after a rescaling $\varphi\varphi'/{\bar\eta}\to \varphi\varphi'$
\beq
Tr\log\left[{\bf 1}+\varphi\varphi'A'-\varphi A\varphi'\right]
+\frac{1}{{\bar\eta}}Tr\left[({\bf 1}+\varphi\varphi'A'-\varphi A\varphi')^{-1}\left(-\varphi H_0\varphi'H_0
+\frac{1}{2}\varphi\varphi'H_0^2+\frac{1}{2}\varphi H_0^2\varphi'\right)
\right]+O({\bar\eta}^{-2})
\ ,
\label{exp01}
\eeq
where $A=iH_0$, $A'=iH_0+4\epsilon$. The first term on the right-hand side gives with Eq. (\ref{leading_sum})
\[
Tr\log\left[{\bf 1}+\varphi\varphi'A'-\varphi A\varphi'\right]
=8\epsilon\sum_\br\varphi_\br\varphi_\br'
\ ,
\]
such that (\ref{exp01}) reads 
\[
8\epsilon\sum_\br\varphi_\br\varphi_\br'
+\frac{1}{{\bar\eta}}\sum_{l=0}^{N-1}
\left\{
Tr\left[\left(\varphi\varphi'A'\right)^l\varphi\varphi'H_0^2\right]
-\sum_{k=0}^l 
Tr\left[\left(\varphi\varphi'A'\right)^{l-k}(-\varphi A\varphi')^k\varphi H_0\varphi'H_0 \right]
\right\}
+O({\bar\eta}^{-2})
\ .
\]
For the expressions in Eq. (\ref{corr00}) we need
\[
\exp\left\{-Tr\log\left[{\bf 1}+4i\eta(\varphi g_+\varphi'-\varphi\varphi' g_-)
-16\eta^2\varphi g_+\varphi'g_-\right]\right\}
\]
which reads with the above expansion
\beq
\exp\left[-8\epsilon\sum_\br\varphi_\br\varphi_\br'\right]
\left[ 1-\frac{1}{{\bar\eta}}\sum_{l=0}^{N-1}
\left\{
Tr\left[\left(\varphi\varphi'A'\right)^l\varphi\varphi'H_0^2\right]
-\sum_{k=0}^l 
Tr\left[\left(\varphi\varphi'A'\right)^{l-k}(-\varphi A\varphi')^k\varphi H_0\varphi'H_0 \right]
\right\}+O({\bar\eta}^{-2})
\right]
\ .
\label{grassmann01}
\eeq
It should be noticed that the diagonal term of $A'$ does not contribute due to the Grassmann
factors. Therefore, we can replace $A'\to A$ here.
Since the Grassmann integation requires a lattice covering of the Grassmann field we need only 
the highest number of $l$ in the sum over $l$. The contribution for $l=N-1$
\[
 Tr\left[\left(\varphi\varphi'A\right)^{N-1}\varphi\varphi'H_0^2\right]
-\sum_{k=0}^{N-1}
Tr\left[\left(\varphi\varphi'A\right)^{N-1-k}(-\varphi A\varphi')^k\varphi H_0\varphi'H_0 \right]
=0
\]
vanishes, as it can be seen by direct inspection of the sum:
\[
Tr\left[\left(\varphi\varphi'A\right)^{N-1}\varphi\varphi'H_0^2\right]
=i^{N-1}Tr\left[\left(H_0\varphi\varphi'\right)^NH_0\right]
\]
and the second term simplifies to same expression,
since the Grassmann variables in the product must have $N$ different coordinates.
The next term with $l=N-2$ in the sum gives a non-vanishing expression
\beq
 Tr\left[\left(\varphi\varphi'A\right)^{N-2}\varphi\varphi'H_0^2\right]
-\sum_{k=0}^{N-2}
Tr\left[\left(\varphi\varphi'A\right)^{N-2-k}(-\varphi A\varphi')^k\varphi H_0\varphi'H_0 \right]
\ .
\label{diff}
\eeq
The second term also reads
\[
i^{N-2}
Tr\left[H_0\left(\varphi\varphi'H_0\right)^{N-2-k}\varphi(-H_0\varphi'\varphi)^kH_0\varphi'\right]
\]
\[
=i^{N-2}\sum_{\br\br'}
Tr_2\left\{
\left[H_0\left(\varphi\varphi'H_0\right)^{N-2-k}\right]_{\br'\br}
\varphi_\br\left[H_0\left(\varphi\varphi'H_0\right)^k\right]_{\br\br'}\varphi'_{\br'}
\right\}
\]
This expression survives the Grassmann integration of Eq. (\ref{grassmann01})
only if $\br'=\br$. Thus, we can write for Eq. (\ref{diff})
\beq
i^{N-2}\left\{
Tr\left[H_0\left(\varphi\varphi'H_0\right)^{N-1}\right]
-\sum_{k=0}^{N-1}\sum_{\br}
Tr_2\left(\left[H_0\left(\varphi\varphi'H_0\right)^{N-2-k}\right]_{\br\br}
\left[\left(\varphi\varphi'H_0\right)^{k+1}\right]_{\br\br}
\right\}\right)
\label{sum11}
\eeq
The last result means that the first expression 
\[
Tr\left[H_0\left(\varphi\varphi'H_0\right)^{N-1}\right]
=\sum_{\br_1,\br_2,...,\br_{N-1}}Tr_2\left(
H_{0;\br_1\br_2}\varphi_{\br_2}\varphi_{\br_2}'H_{0;\br_2\br_3}
\cdots \varphi_{\br_{N-1}}\varphi_{\br_{N-1}}'H_{0;\br_{N-1}\br_1}\right)
\]
is nonzero only when $\br_1$ is different from all the other sites $\{\br_j\}$ ($j=2,..., N-1$).
This can be inserted in the Grassmann integral with the expression (\ref{grassmann01})
by borrowing a factor $-8\epsilon\varphi_{\br_1}\varphi_{\br_1}'$ from the first exponential 
factory to obtain Eq. (\ref{det11}).

For the product
$
\varphi_{\bar\br}\varphi_{{\bar\br}'}'
\det\left({\bf 1}+\varphi \varphi'-\varphi h\varphi'h^\dagger\right)^{-1}
$
we get
\[
\varphi_{\bar\br}\varphi_{{\bar\br}'}'
\exp\left[-8\epsilon\sum_\br\varphi_\br\varphi_\br'\right]
\]
\[
\left[ 1-\frac{1}{{\bar\eta}}\sum_{l=0}^{N-1}
\left\{
Tr\left[\left(\varphi\varphi'A'\right)^l\varphi\varphi'H_0^2\right]
-\sum_{k=0}^l 
Tr\left[\left(\varphi\varphi'A'\right)^{l-k}(-\varphi A\varphi')^k\varphi H_0\varphi'H_0 \right]
\right\}+O({\bar\eta}^{-2})
\right]
\]
and with $\br\ne\br'$
\beq
=\frac{i^{N-2}}{{\bar\eta}}\varphi_{\bar\br}\varphi_{{\bar\br}'}'
\sum_{k=0}^{N-2} 
Tr\left[\left(\varphi\varphi' H_0\right)^{N-2-k}(-\varphi H_0\varphi')^k\varphi H_0\varphi'H_0 \right]
+O({\bar\eta}^{-2})
\ .
\label{sum12}
\eeq
Moreover, we have
\[
Tr\left[\left(\varphi\varphi' H_0\right)^{N-2-k}(-\varphi H_0\varphi')^k\varphi H_0\varphi'H_0 \right]
=Tr\left[\left(H_0\varphi\varphi'\right)^{N-2-k}H_0\varphi (H_0\varphi\varphi')^k H_0\varphi'\right]
\ ,
\]
which also reads
\[
\sum_{\br_1,\br_2,...,\br_{N-1}}Tr_2\Big(
\left[H_{0;\br_1\br_2}\varphi_{\br_2}\varphi_{\br_2}'\cdots H_{0;\br_{N-3-k}\br_{N-2-k}}\varphi_{\br_{N-2-k}}\varphi_{\br_{N-2-k}}'\right]
H_{0;\br_{N-2-k}\br_{N-1-k}}\varphi_{\br_{N-1-k}}
\]
\[
\left[H_{0;\br_{N-1-k}\br_{N-k}}\varphi_{\br_{N-k}}\varphi_{\br_{N-k}}'\cdots H_{0;\br_{N-1}\br_{N}}\varphi_{\br_{N}}\varphi_{\br_{N}}'\right]
H_{0;\br_{N}\br_1}\varphi_{\br_1}'
\Big)
\ .
\]
When we multiply this expression with $\varphi_{\bar\br}\varphi_{{\bar\br}'}'$ it gives a nonzero result only if $\br_{N-1-k}={\bar\br}'$ and
$\br_1={\bar\br}$. Thus we get for the expression in (\ref{sum12})
\beq
-\frac{i^{N-2}}{{\bar\eta}}\sum_{k=0}^{N-2} 
Tr_2\left[\left(H_0\varphi\varphi'\right)^{N-1-k}_{{\bar\br}{\bar\br}'}(H_0\varphi\varphi')^{k+1}_{{\bar\br}'{\bar\br}}\right]
+O({\bar\eta}^{-2})
\ .
\label{sum13}
\eeq

\end{document}